\documentclass[12pt]{article}
\usepackage{graphicx}
\usepackage{amssymb}
\usepackage{amsfonts}

\title{Differential Equations and Applications to COVID-19 }
\author{Mitonsou Tierry Hounkonnou$^{1, \star}$ and Laure Gouba$^{1, 2,\diamond}$ \\
$^{1}${\em
African Institute for Mathematical Science (AIMS-S\'en\'egal)} \\
{\em Km 2 routes de Joal (Centre IRD Mbour) B. P. 1418 Mbour, S\'en\'egal.}\\
$^{\star}${\em Email: tierryhma@gmail.com}\\
$^{2}${\em
The Abdus Salam International Centre for Theoretical Physics (ICTP)}\\
{\em Strada Costiera 11, I-34151 Trieste Italy.}\\
$^{\diamond}$ {\em Email: laure.gouba@gmail.com}}

\begin{document}

\maketitle

\begin{abstract}
\noindent
This paper focuses on the application of the Verhulst logistic equation to model in retrospective the total COVID-19 cases in Senegal during the period from April 2022 to April 2023. Our predictions for April 2023 are compared with the real COVID-19 data for April 2023 to assess the accuracy of the model. The data analysis is conducted using Python programming language, which allows for efficient data processing and prediction generation. 
\end{abstract}

\section{Introduction}

The COVID-19 outbreak, caused by the SARS-CoV-2 virus, was initially identified in Wuhan,
China, in December 2019 \cite{jernigan}. The virus rapidly spread across multiple countries, leading to a significant number of global fatalities.  The outbreak was declared a pandemic on March 2, 2020 \cite{anjorin}. As the pandemic is growing , they have been concerns about finding treatments, vaccines, and although the lockdowns, many researchers develop algorithms to predicts the new cases, some researchers use well known mathematical methods, such as a mathematical logistic growth model.

Senegal reported its first case of COVID-19 on March 1, 2020 and its COVID-19 data are known from the early stage of the pandemic to nowadays. We are interested in studying the case of Senegal over a period of one year in a retrospective way, from April 2022 to April 2023.  We fit the solution of the Verhulst logistic equation to the data from April 2022 to April 2023 using Python. Since the Data are furnished for April 2023 we compare the predictions and the real data.

Our work is organized as follows. 
Section \ref{sec2}, is about COVID-19. In section \ref{sec3} , we present the Verhulst logistic equation. Section \ref{sec4} is about the data
analysis. Through this analysis, we seek to uncover the trends and patterns of COVID-19 cases over a specific time frame, specifically from April 2022 to April 2023. The conclusion is given in section \ref{sec5}.

\section{COVID-19}\label{sec2}

COVID-19, caused by the severe acute respiratory syndrome coronavirus-2 (SARS-CoV-2)\cite{niu}, emerged unexpectedly in 2019, profoundly impacting human life \cite{kalifa}. The initial outbreak occurred in Wuhan City, China \cite{spiteri}, and rapidly spread worldwide, including the United States, Europe, Asia, and other continents. On January $30, 2020$, the World Health Organization (WHO) declared it a global public health emergency, and on March $2, 2020$, it was declared a pandemic. COVID-19 is the most disruptive infectious disease
the world has experienced, ranking second in documented cases after human immunodeficiency virus (HIV). Africa’s response to COVID-19 has exhibited a distinct pattern compared to other regions, with
studies indicating a slower rate of exponential growth in the disease burden \cite{bouba}.
As of November 2022, Africa accounted for a relatively small proportion of global COVID-19 cases
and fatalities \cite{organization}. To mitigate the spread of the virus, African nations have
implemented various non-pharmaceutical interventions (NPIs) such as social distancing, mask-wearing, and testing  \cite{mendez}. These measures have played a significant role in reducing the impact of the pandemic, although the emergence of new SARS-CoV-2 variants has presented challenges \cite{Wang}. Vaccination has become a critical strategy in the
battle against the virus and its variants \cite{shattock}. On March 1, 2020, Senegal confirmed its first case of COVID-19 \cite{diarra}. In response, the country implemented a range of measures to curb the virus, including travel restrictions,
lockdowns, hygiene promotion, and widespread testing. Senegal initiated its vaccination campaign
in February 2021, initially relying on the COVAX facility for vaccine supply. The AstraZeneca
vaccine was prioritized for healthcare workers, elderly individuals, and those with underlying health
conditions. Bilateral agreements and donations also provided additional doses, including the
Sinopharm vaccine from China \cite{suzuki}. The vaccination efforts aimed to cover
the entire population, particularly focusing on high-risk and vulnerable groups.
Various preventive measures were implemented by governments and stakeholders, including lock-
downs, restrictions on gatherings, hygiene promotion, social distancing, and mask mandates in
public places. Efforts were made to equip treatment facilities, recruit healthcare professionals, and
incentivize frontline workers\cite{gollwitzer}. Despite these measures, the virus continues
to spread, with surges observed in Senegal and other countries worldwide. As of February 24,
2021, Africa had reported 3,872,085 cases and 102,286 deaths \cite{iddrisu}. Ongoing
statistics indicate a clear upward trend in cases, highlighting the persistent challenges posed by
the virus.

COVID-19 stands for ``Coronavirus Disease 2019”. Let’s break down the meaning of each component
\begin{itemize}
\item ``CO” denotes ``corona,” representing the family of viruses that includes SARS-CoV-2, the
virus responsible for COVID-19. Coronaviruses encompass a wide range of viruses, causing
various illnesses from common colds to severe diseases.
\item 
`` VI ” stands for `` virus,” a tiny infectious agent that can only reproduce within the cells of
a living organism. In the case of COVID-19, the virus is known as SARS-CoV-2.
\item 
``D” represents ``disease,” referring to the illness caused by the SARS-CoV-2 virus. COVID-
19 specifically affects the respiratory system, leading to a spectrum of symptoms, ranging
from mild to severe.
\item ``19” signifies the year 2019 when the disease was first identified. The initial cases of
COVID-19 were reported in Wuhan, Hubei Province, China, in December 2019.
\end{itemize}
Hence, COVID-19 represents the particular illness triggered by the SARS-CoV-2 virus, originating
in 2019 and subsequently disseminating worldwide as a pandemic.

\begin{figure}[h!]
\centering
\includegraphics[width=12cm]{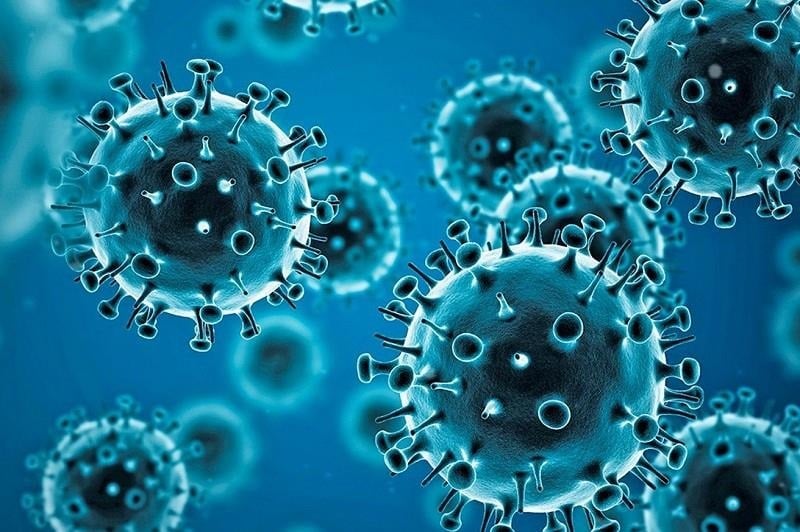}
\caption{Coronavirus}
\end{figure}

Typical symptoms of COVID-19, which can manifest within 2-14 days after exposure to the SARS-CoV-2 virus, range from mild to severe and commonly include:
\begin{itemize}
\item Fever: An elevated body temperature above 
$38^\circ $ C ( $100.4^\circ $ F) is a common symptom of COVID-19.
\item Cough: A dry cough is a frequent symptom, which may persist and become more severe
over time.
\item Shortness of breath: Some individuals may experience difficulty breathing or shortness of
breath, particularly in severe cases.
\item Fatigue: Feeling tired or experiencing extreme exhaustion is a common symptom reported
by COVID-19 patients.
\item Muscle or body aches: Muscle pain, body aches, or general discomfort can occur with
COVID-19.
\item Headache: Many individuals with COVID-19 may experience headaches, which can range
from mild to severe.
\item Sore throat: Soreness or irritation in the throat may be present, often accompanied by a
dry cough.
\item Loss of taste or smell: An altered sense of taste (dysgeusia) or loss of smell (anosmia) has
been reported by some individuals infected with COVID-19.
\item Congestion or runny nose: While less common, nasal congestion or a runny nose can occur
in some cases.
\item Gastrointestinal symptoms: Some individuals may experience symptoms like nausea, vom-
iting, or diarrhea.
\end{itemize}

It’s important to note that not everyone infected with COVID-19 will experience symptoms, and some individuals may have mild or no symptoms at all. Additionally, symptoms can vary widely between individuals, and some symptoms may overlap with other respiratory illnesses. If you experience any concerning symptoms or have been in close contact with someone who tested
positive for COVID-19, it is recommended to consult with a healthcare professional and follow
local health guidelines for testing and care.

\subsection{How is COVID-19 spread?}

Transmission of COVID-19 primarily occurs through respiratory droplets that are emitted when
infected individuals cough, sneeze, talk, or breathe. The virus, SARS-CoV-2, can be transmitted through the following routes:
\begin{itemize}
\item Avoiding direct physical contact, such as touching one another.
\item Indirect contact refers to the transmission of the virus through contact with contaminated
objects or surfaces. When individuals infected with the virus sneeze, cough, or touch surfaces, they can leave behind infected droplets on these surfaces. The virus can remain viable on these surfaces for varying periods, ranging from several hours to a few days.
\item  Close contact refers to the situation where individuals are in proximity to an infected person
and come into contact with the respiratory droplets emitted from their nose and mouth.
\item Transmission through the air is possible in crowded indoor environments with limited 
ventilation. Small droplets containing the virus can travel greater distances under these conditions.
Therefore, meeting people outdoors is considered safer than indoors, even if you maintain
a distance of more than 2 meters.
\item Transmission can also occur through contaminated surfaces. When an individual who is
infected with the virus sneezes or coughs, droplets containing the virus can land on nearby
surfaces. If you touch these surfaces and then touch your eyes, nose, or mouth, you can
potentially become infected as well.
\item Inhaling air in close proximity to an infected person who is exhaling droplets containing the
virus can also lead to transmission.
\item Certain medical procedures can also facilitate the transmission of the virus.
\item Aerosol transmission refers to the spread of infected droplets that can remain suspended
in the air for extended periods, particularly in indoor environments with limited fresh air
circulation.
\item  Fan and air-conditioner which circulated the infected droplets (especially if they 
recirculate the air).
\end{itemize}

\subsection{Treatment}

At present, there is no approved treatment specifically for COVID-19. The management of
the disease depends on its severity, ranging from supportive care for mild cases to more intensive
interventions for severe cases. The primary focus is on providing symptomatic and supportive care,
including monitoring vital signs, ensuring adequate oxygen levels, maintaining blood pressure, and
addressing complications like secondary infections or organ failure. Given the potential fatality of COVID-19, numerous investigational treatments are currently being studied:
\begin{itemize}
\item Symptom management: Many COVID-19 cases involve mild symptoms that can be managed at home.
\item Hospital care: For severe cases, hospitalization may be necessary. This can involve providing supplemental oxygen therapy to maintain adequate oxygen levels, administering intravenous
fluids, and monitoring vital signs closely.
\item Corticosteroids: In severe cases, corticosteroids like dexamethasone are used to reduce
inflammation and suppress an overactive immune response. They have been shown to improve outcomes and reduce mortality rates in hospitalized patients requiring oxygen therapy or mechanical ventilation.
\item Immune-based therapies: Monoclonal antibodies, convalescent plasma, and other immune-
based therapies are used in specific cases to help boost the immune response and neutralize the virus. These treatments are typically reserved for individuals at high risk of severe disease or those with moderate symptoms.
\item They have been shown to improve outcomes and reduce mortality rates in hospitalized patients requiring oxygen therapy or mechanical ventilation.
\item In severe cases of COVID-19 with significantly low oxygen levels, respiratory support such as
mechanical ventilation or extracorporeal membrane oxygenation (ECMO) may be required to provide adequate oxygen therapy.
\end{itemize}

It’s crucial to acknowledge that the treatment approach for COVID-19 is continuously evolving
with ongoing research and clinical trials. Healthcare professionals should be consulted to determine the most suitable treatment options based on factors such as patient characteristics, disease severity, and the latest insights.

\subsection{Prevention of rapid spread and disease progression of COVID-19}

In order to prevent infection and promote effective immune responses to the virus, it is crucial to
ensure adequate indoor air circulation, practice proper cough etiquette, and maintain good hand
hygiene. Additionally, implementing social and physical distancing measures among individuals
is essential to prevent overwhelming the healthcare system. It may be necessary to temporarily
accommodate an increased number of patients, as long as it remains within the capacity of
the current healthcare infrastructure. These strategies align with appropriate healthcare policies, such as the pursuit of herd immunity as observed in Sweden (Giesecke, 2020). By implementing
these measures, there is a reasonable expectation that we can successfully navigate the projected
second wave anticipated in the northern hemisphere during the late fall and winter seasons (Trilla
et al., 2008). Here are some key strategies:
\begin{itemize}
\item Vaccination: Getting vaccinated against COVID-19 is highly effective in preventing severe
illness, hospitalization, and mortality. Vaccines have been developed and approved for emergency use in many countries. Following the recommended vaccination schedule and receiving booster shots, if necessary, can provide significant protection against the virus.
\item Practice proper hygiene by consistently washing your hands with soap and water for at least 20 seconds, particularly after being in public spaces or touching surfaces. If soap and water are not accessible, utilize hand sanitizer that contains at least 60 $\%$  alcohol. Avoid touching your face, especially your eyes, nose, and mouth, as this can lower the chances of viral transmission.
\item  Use masks: When in indoor public spaces or crowded outdoor settings, wear a mask that
covers your nose and mouth. Masks play a crucial role in preventing the spread of respiratory
droplets that may contain the virus. Adhere to local guidelines and regulations regarding
mask usage.
\item Maintain physical distance: Practice social distancing by keeping a distance of at least
1 meter (3 feets) from others, especially in situations where physical distancing may be
challenging, such as crowded places or close-contact settings.
\item Ventilation and air circulation: Ensure proper ventilation in indoor spaces by opening win-
dows or using mechanical ventilation systems. Good airflow can help dilute and remove
viral particles from the air, reducing the risk of transmission.
\item Avoid large gatherings: Limit or avoid gatherings, especially in enclosed spaces, where the
risk of transmission is higher. Instead, opt for virtual meetings or outdoor activities that
allow for better physical distancing.
\item Stay home if unwell: If you have symptoms related to COVID-19, such as fever, cough,
or breathing difficulties, it is important to stay at home and consult with a healthcare
professional. Follow local guidelines on testing, self-isolation, and quarantine protocols.
\item Follow public health guidelines: Stay informed about the latest guidance and recommen-
dations provided by local health authorities, including travel advisories, testing protocols,
and quarantine requirements. Adhere to these guidelines to protect yourself and others.
\item Enhanced cleaning and disinfection: Regularly clean and disinfect frequently touched surfaces, particularly in areas with high foot traffic. Use approved disinfectants that are
effective against COVID-19.
\item Support contact tracing efforts: Cooperate with contact tracers if you test positive for
COVID-19 or have been in close contact with someone who has. Providing accurate information can help identify potential transmission chains and prevent further spread.
\end{itemize}
These preventive measures work in conjunction with one another to reduce the spread of the virus. It’s important to adapt and follow the specific guidelines and recommendations provided by health authorities in your region, as they may vary depending on the local epidemiological situation and vaccination rates.

\section{Verhulst logistic equation}\label{sec3}

The Verlhulst differential equation, known as the Verlust logistic equation, is a first order differential equation due to Pierre Fran\c{c}ois Verlhust, and of the form 
\begin{equation}\label{ver1}
\frac{dP}{dt} = aP \left( 1- \frac{P}{E}\right), 
\end{equation}
where $P(t)$ represents the population size, $a$ is the population rate which is the Malthusian parameter and $E$ is the carrying capacity of the Environment \cite{andriani}. The story of the equation (\ref{ver1}) rooted from the work of Adolphe Quetelet in 1835 on `` Man and the Development of his faculties", where he introduced the concept of obstacles limiting long-term geometric population growth. Queterlet proposed that these obstacles create resistance proportional to the square of the population growth rate, which inpires Verlhust to propose a differential equation,  (\ref{ver1}), to model population growth as an alternative approach. When the population $P(t)$ in the equation (\ref{ver1}) is significantly smaller than $E$, an approximate equation 
\begin{equation}
\frac{dP}{dt}
\end{equation}
can be used, resulting in exponential growth 
\begin{equation}
P(t) = P(0) e^{at}.
\end{equation}
When $P(t)$ is negative, the growth rate would become negative. 
In $1847$, Verlhust introduced a modified differential equation 
\begin{equation}\label{ver2}
\frac{dP}{dt} = a (P-b)\left( 1 - \frac{P-b}{E}\right).
\end{equation}
 Despite Verlhust's initial indecision regarding model equations, the logistic equation reemerged independently several decades later from various sources. The term ``carrying capacity " represents the maximum population, denoted as E, in the logistic equation. The Verlhust equation (\ref{ver1}) can be solved in few steps
 \begin{itemize}
 \item First step: dividing the equation (\ref{ver1}) by $p^2$, one obtains 
 \begin{equation}
 \frac{d}{dt}\left[ \frac{1}{p}\right] = a\left(
 \frac{1}{E} - \frac{1}{P} \right).
 \end{equation}
 \item Defining $x(t) = \frac{1}{P(t)}$, we have 
 \begin{equation}\label{ver3}
 \frac{d}{dt} x(t)  = - a \left( x(t) - \frac{1}{E}\right).
 \end{equation}
 \end{itemize}
The solution of the equation (\ref{ver3}) is given by 
\begin{equation}
x(t) = \left(\frac{1}{p(0)} - \frac{1}{E} \right) e^{-at} + \frac{1}{E}.
\end{equation}
Since $x(t) = \frac{1}{p(t)}$, the solution $P(t)$ of the equation (\ref{ver1}) is given by $P(t) = [x(t)]^{-1}$, which is explicitly 
\begin{equation}
P(t) = \frac{E P_0}{e^{-at}[E -P_o]+ P_o}.
\end{equation}

\section{Data analysis}\label{sec4}

This study focuses on Senegal, a country located in West Africa, which coordinates approximatively between $12^\circ$ N and $17^\circ$ N latitude $17^\circ$ W and $16^\circ$ W longitude. Senegal has  land area of around $196, 712$ square kilometers and an estimated population of about $18, 015, 360$ people as of May 22, 2023. Senegal is border with Mauritania, Mali, Guinea, Guinea Bissau, and the Gambia, with the Atlantic Ocean forming its western border. It has a tropical climate with distinct wet and dry seasons. The average daily temperature ranges from $25^\circ$ C to $35^\circ$ C, and the relative humidity varies between 50 percent and 90 percent. The rainy season occurs from June to October, while the dry season spans from November to May. 
The Verhulst logistic model was employed to fit the time series of total COVID-19 cases. The
model’s parameters represent the maximum number of confirmed cases (E) and the growth rate
(a) of the curve. The calculations were performed using Python.

\begin{figure}[h!]
\centering
\includegraphics[width=12cm]
{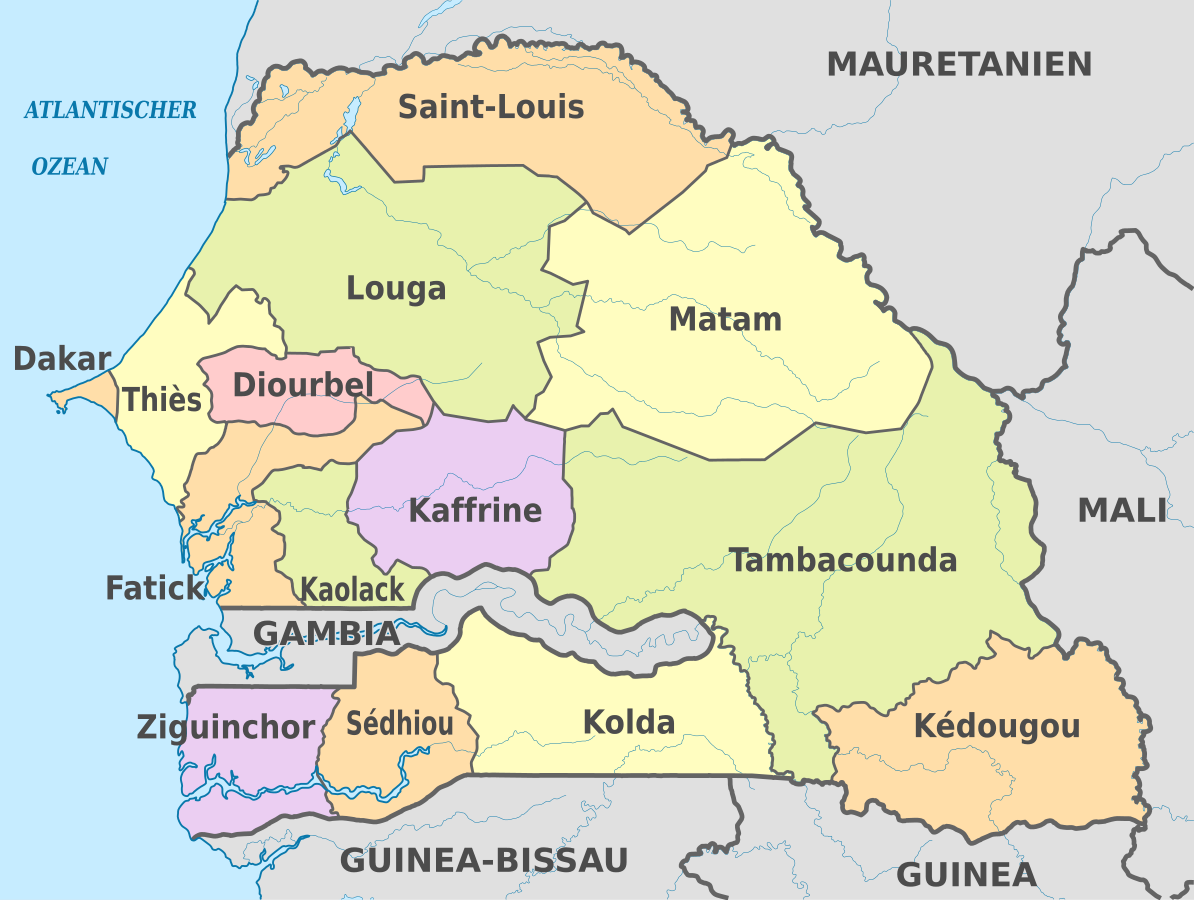}
\caption{Regional Map of Senegal: Source: https://www.worldatlas.com/maps/senegal}
\end{figure}

\begin{figure}[h!]
    \centering
    \includegraphics[width=12cm]{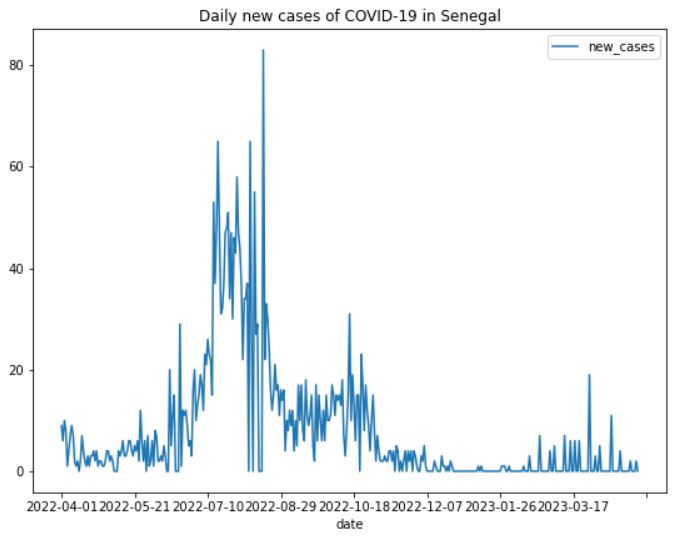}
    \caption{Daily New cases of CoVID-19}\label{fig3}
\end{figure}

Data from COVID-19 are available. Location, date, total cases, new cases, total deaths, and
new deaths are the six variables that make up the COVID-19 data that was used for the study.
The information covered the time period from April 1, 2022, through April 30, 2023. There
are 395 observations for each variable in the daily COVID-19 data, which were taken from
(https://github.com/owid/covid-19-data/tree/master/public/data, ).
\newpage 
The Figure \ref{fig3} displays the trend of the number of new cases and fatalities each day, as well as the
number of cases and deaths overall, for COVID-19 in Senegal through time. It offers information
on the rate of infection and aides in comprehending the peaks and valleys in the quantity of
new cases. The combination of these four subplots offers a complete picture of the COVID-19
situation in Senegal.

\begin{figure}[h!]
    \centering
    \includegraphics[width=12cm]{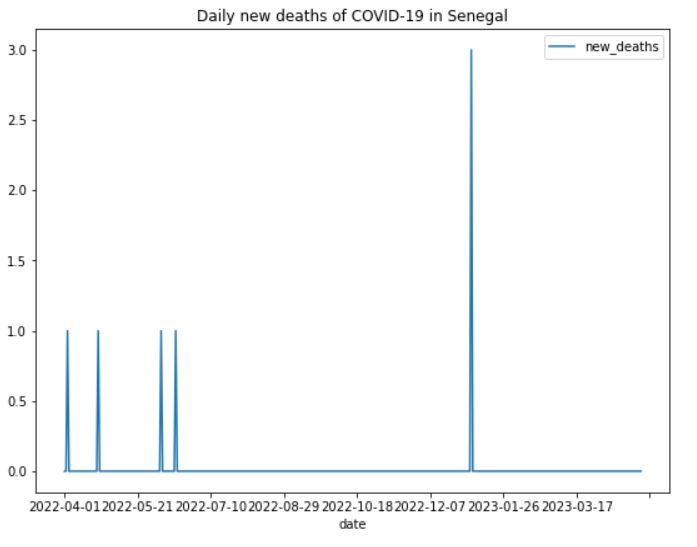}
    \caption{Daily new deaths of COVID-19}\label{fig4}
\end{figure}

\begin{figure}[h!]
    \centering
    \includegraphics[width=12cm]{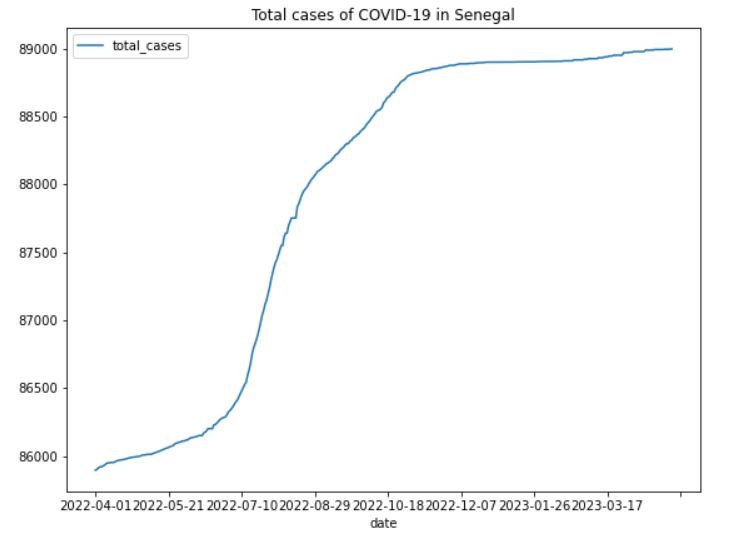}
    \caption{Total cases of COVID-19 in Senegal}\label{fig5}
\end{figure}

\begin{figure}[h!]
    \centering
    \includegraphics[width=12cm]{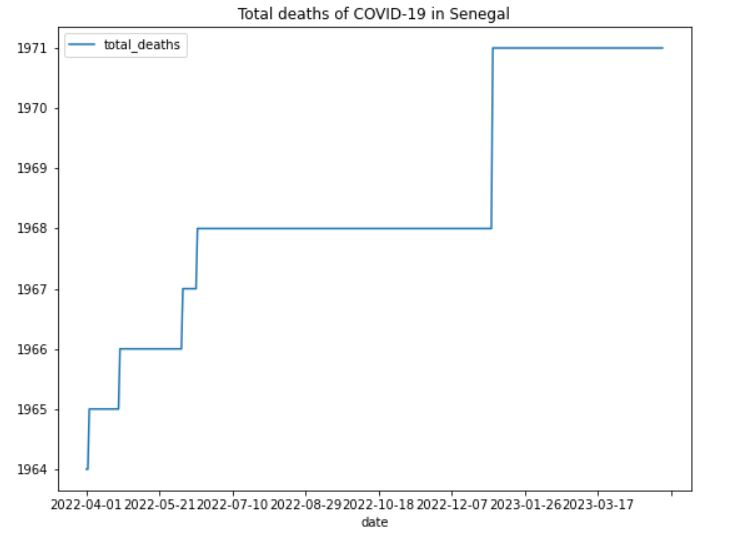}
    \caption{Total deaths of COVID-19 in Senegal}\label{fig6}
\end{figure}

\subsection{Data analysis over the period April 2022 to April 2023}

\subsubsection{ Descriptive statistics and distribution of the considered data set }
The table \ref{tab41} below provides an overview of the descriptive statistics for the COVID-19 cases
used in the study.

\begin{table}[h!]\label{ab41}
 \begin{center}
  \begin{tabular}{rrrrrrr}
   \hline variables & mean & sd & median & min & max & se \\
   \hline Total cases & 87923.12 & 1192.00 & 88601.00 & 85895.0 & 88997.00 & 59.90 \\
   New cases & 7.87 & 12.48 & 3.00 & 0.00 & 83.00 & 0.62 \\
   Total deaths & 1968.46 & 1.86 & 1968.00 & 1964.00 & 1971.00 & 0.09 \\
   New deaths & 0.01 &0.18 & 0.00 & 0.00 & 3.00 & 0.01 \\
   \hline
  \end{tabular}
  \vspace{0.1cm}
  \caption{Descriptive statistics of COVID-19 cases.} 
  \label{tab:descriptive}
 \end{center}
\end{table}

The average number of new COVID-19 cases per day was 7.87, with a standard deviation of
12.48, while the average daily death rate was 0, with a standard deviation of 0.18. The synopsis
also revealed that the distribution of COVID-19 deaths and newly observed daily cases is positively
biased. in Senegal, the number of total case is 88997.

\begin{figure}[h!]\label{dataset}
\centering
\includegraphics[width=12cm]
{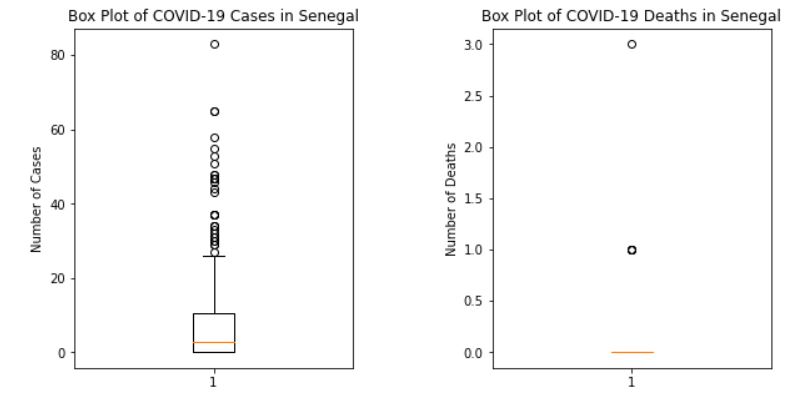}
\caption{Plots of the COVID-19 data set}
\end{figure}

\subsubsection{Results and discussions}

Mathematical models of various complexities are employed to understand and predict the spread
of epidemics. The Verhulst logistic equation, a nonlinear first-order ordinary differential equation
(ODE), is historically the first model used for this purpose \cite{verhulst}. It is commonly
utilized to describe population growth and advertising performance. In the context of COVID-19,
this model has been applied in previous studies \cite{giorgio}. The analysis of
COVID-19 data suggests different growth patterns in different regions, with exponential growth
observed in some countries and power law growth in others \cite{natalia}. To capture
these dynamics, the generalized logistic equation is utilized \cite{blumberg, fabio}. Despite its mathematical simplicity, the logistic equation provides valuable insights.
Various statistical tests were employed to examine the COVID-19 responses in Senegal.
To evaluate the randomness of COVID-19 responses in Senegal, the Ljung-Box test was conducted
using Python software. The test results indicated that the p-values (0.06864417 for COVID-19
responses and 0.7643388 for deaths) were both greater than the significance level of 5 percent. 
Therefore, we cannot reject the null hypothesis that the responses and deaths are random, suggesting
no autocorrelation in Senegal.
The Keenan’s one-degree test for non-linearity was employed to analyze the autoregressive nature
of the daily confirmed cases and daily deaths time series in Senegal. Python statistical software
was used for this purpose. The test yielded a p-value of 0.00 for the daily confirmed cases, which
is below the significance level of 0.05. Hence, we reject the null hypothesis and conclude that
the daily confirmed cases in Senegal do not exhibit an autoregressive process. Similar conclusions were drawn for the number of death cases, as the p-value 
$(0.00)$ was also below the significance level.

\begin{table}[h!]
  \centering
   \caption{Ljung-Box test on COVID-19: Total Cases in Senegal} \label{tab:cases}
  \begin{tabular}{rr}
   \hline 
   & Total Cases of COVID-19 \\
   \hline 
   Statistics & 17.26781 \\
   P-value & 0.06864417 \\
   \hline
  \end{tabular}
\end{table}
 
 \begin{table}[h!]
  \centering
  \caption{Ljung-Box test on COVID-19: Total Deaths in Senegal}\label{tab:deaths}
  \begin{tabular}{rr}
   \hline 
   & Total Deaths of COVID-19 \\
   \hline 
   Statistics & 6.580794 \\
   P-value & 0.7643388 \\
   \hline
  \end{tabular}
\end{table}

\begin{table}[h!]
  \centering
   \caption{Keenan's one-degree test for non-linearity on COVID-19: New Cases in Senegal} \label{tab:cases}
  \begin{tabular}{rr}
   \hline 
   & New Cases of COVID-19 \\
   \hline 
   Statistics & -15.66869 \\
   P-value & 0.00 \\
   \hline
  \end{tabular}
  \end{table}
  
 \begin{table}[h!]
  \centering
  \caption{Keenan's one-degree test for non-linearity on COVID-19: New Deaths in Senegal} \label{tab:deaths}
  \begin{tabular}{rr}
   \hline 
   & New Deaths of COVID-19 \\
   \hline 
   Statistics & -30.73934 \\
   P-value & 0.00 \\
   \hline
  \end{tabular}
\end{table}

\subsection{Trends over the period April 2022 - April 2023}

In summary, the logistic model effectively captures the progression of COVID-19 cases over time,
despite its simplicity.
The logistic function defined as ``logistic(t, a, K)" solves the standard logistic differential equation
given by the equation (\ref{ver1}). This equation is commonly used to model population growth or the
spread of a phenomenon over time. The figure 4.8 shows the plots of the real COVID-19 data
and the logistic curve.

\begin{figure}[h!]
\centering
\includegraphics[width=12cm]{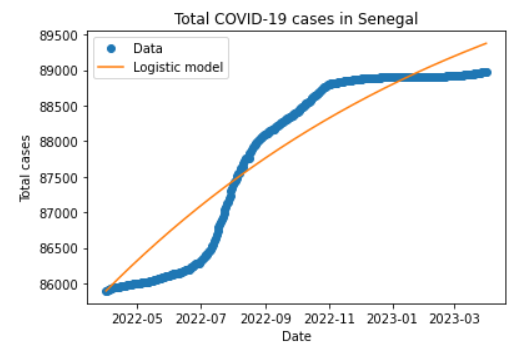}
\caption{Daily New cases of CoVID-19}\label{fig7}
\end{figure}
 
\begin{figure}[h!]
\centering
\includegraphics[width=12cm]{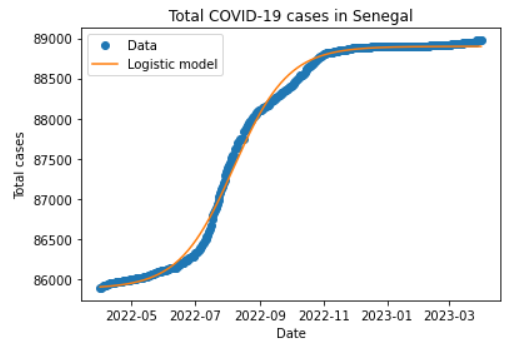}
\caption{Daily New cases of CoVID-19}\label{fig8}
\end{figure}

 From the figure \ref{fig7} we observed that the logistic function does not fit the data well. This is likely due to the fact that COVID-19 cases do not start from zero; there is an existing baseline level.
Looking for a best fit of the data, we add an additional constant term `` b ” to the logistic function denoted as ``logistic offset(t, a, K, b, N0)". Its solves a modified version of the logistic equation (\ref{ver2})

\begin{equation}
\frac{dP}{dt} = a (P-b)(1 - \frac{P-b}{E}),
\end{equation}
where the term `` b" introduces the additional constant and allows for an offset or baseline value in the logistic function. This modified equation incorporates an offset or baseline value, represented
by the constant `` b ". It is particularly useful when the phenomenon being modeled has a baseline
or an initial value that is different from zero. 
The figure \ref{fig8} shows the plots of the real COVID-19 data and the logistic curve with the additional constant term.
From the figure \ref{fig8}, we observed that the logistic function with the additional constant term,
which incorporates the offset or baseline value, provides a good fit to the COVID-19 data. This
modification allows for a more accurate representation of the initial conditions and external
influences affecting the COVID-19 spread in Senegal.
Based on the figures ( \ref{fig7}, \ref{fig8} ), it can be inferred that incorporating an additional
constant term in the logistic function provides a better fit for modeling COVID-19 data in Senegal
compared to the logistic function without the constant term. The inclusion of the constant ’b’
enhances the representation of initial conditions and baseline, resulting in improved alignment
with the observed data.

\subsection{Comparison/discussion with predicted results from April 2022 to 2023 }

To assess the accuracy and reliability of the logistic model, we compared the predicted COVID-19
cases with the actual data. Performance metrics like mean squared error (MSE) and coefficient
of determination (R-squared) were used in this study to evaluate the model’s goodness-of-fit and predictive ability.
\begin{enumerate}
\item The average of the squared differences between predicted and actual values is calculated
using the mean squared error (MSE). It gives an indication of the overall difference between
the model’s predictions and reality. The squared discrepancies between each predicted value
and its matching actual value are averaged to produce the MSE.
\begin{equation}
\textrm{MSE} = \frac{1}{n}\sum \left( 
y_{\textrm{true}} - y_\textrm{pred}\right)^2
\end{equation}
A smaller MSE value indicates superior model performance, indicating minimal discrepancies
between predicted and actual values. A MSE of zero represents a perfect fit of the model
to the data. It is important to consider that outliers can have a significant impact on the
overall score of MSE, as their squared differences exert a greater influence.
\item 
R-squared, or the coefficient of determination, indicates the percentage of variance in the
dependent variable that can be accounted for by the independent variable(s) in the model.
It evaluates the degree to which the model fits the data.
R-squared is calculated as:
\begin{equation}
R-\textrm{squared} = 1 - \frac{SS_{\textrm{residual}}}{SS_{\textrm{total}}}
\end{equation}
\end{enumerate}
where: SS residual is the sum of squared residuals (differences between the actual and predicted
values) and SS total is the total sum of squares, which measures the total variance of the
dependent variable.
The evaluation of the logistic model’s performance was conducted on the COVID-19 data for
Senegal from April 2022 to April 2023.
The table \ref{tab:evaluation_metrics} displays the result of the differente performance metric value on the cases of COVID-19 in Senegal. 
From the table \ref{tab:evaluation_metrics}, it can be observed that the mean squared error (MSE)is equal to 152054.68 and the coefficient of determination 
(R-squared) is 0.89.
As Rsquared value closer to 1, it indicates a stronger relationship and better predictive perfor-
mance of the model.

\begin{table}[h]
 \centering
 \begin{tabular}{|c|c|}
  \hline
  \textbf{Evaluation Metric} & \textbf{Value} \\
  \hline
  Mean Squared Error (MSE) & 152054.68 \\
  \hline
  Coefficient of Determination (R-squared) & 0.89 \\
  \hline
 \end{tabular}
 \caption{Evaluation Metrics for Logistic Model Predictions}
 \label{tab:evaluation_metrics}
\end{table}

Finally, we calculated the Mean Absolute Percentage Error (MAPE) to assess the model’s accu-
racy. The average percentage difference between the expected and actual values is quantified.
The following equation can be used to determine MAPE
\begin{equation}
\textrm{MAPE} = \frac{1}{n}\times \sum \vert \frac{\textrm{Actual} - \textrm{Predicted}}{\textrm{Actual}} \times 100\,,
\end{equation}
where n is the number of data points, Actual represents the actual values and Predicted represents the predicted values. MAPE is expressed as a percentage, indicating the average percentage error in the predictions.
It provides a relative measure of the accuracy, allowing comparison across different datasets and
models.
In this thesis, the Mean Absolute Percentage Error (MAPE) is calculated using Python to evaluate
the accuracy of the predictions. The MAPE value obtained is $0.0019$ which is close to $0 \%$ . This
result indicates a very accurate prediction, meaning that the predicted values are very close to
the actual values (the reality of April 2023).

\section{Conclusion}\label{sec5}

The pandemic has significantly impacted the
global population, leading to widespread health, social, and economic consequences. Indeed, 
the social and economic consequences of COVID-19 has been widespread and significant, affecting individuals, communities, and economies around the world. The pandemic has caused significant disruptions to daily life, including school closures, job losses, and reduced social interactions. The impact has been particularly pronounced for vulnerable populations, including those
with pre-existing health conditions, low-income households, and marginalized communities. In
terms of the economy, COVID-19 has had a significant impact on businesses, with many small businesses closing permanently due to decreased revenue and increased costs associated with adapting to the pandemic. The pandemic has also led to disruptions in global supply chains, reducing the availability of goods and increasing prices for certain products.
The COVID-19 pandemic has highlighted and worsened pre-existing social and economic disparities. Vulnerable communities have been disproportionately impacted, experiencing higher rates of illness, death, job losses, reduced access to support systems, and decreased educational opportunities. Additional key areas affected by the pandemic include:

Mental Health Impact: The COVID-19 pandemic has had a profound effect on mental well-being.
Social isolation, fear, anxiety, and grief associated with the virus have contributed to a rise in
mental health disorders. The strain on healthcare systems has also limited access to mental health
services, exacerbating the mental health crisis.
The COVID-19 pandemic has underscored the significance of social cohesion and community
resilience. Many communities have come together to support each other through mutual aid
networks, volunteering, and community initiatives. This crisis has emphasized the significance of
collective action. 

Unemployment and Income Loss: The economic impact of COVID-19 has resulted in a significant
rise in unemployment rates. Many people have lost their jobs or experienced reduced working
hours, leading to income loss and financial instability. This has disproportionately affected vulnerable populations, including low-income workers, informal sector workers, and those in precarious
employment.

Poverty and Inequality: The pandemic has exacerbated existing social inequalities and pushed
many people into poverty. The loss of livelihoods and reduced access to basic services have left
vulnerable populations struggling to meet their basic needs. Women, children, the elderly, and
marginalized communities have been particularly affected by the socio-economic consequences of
the pandemic.

In this work, we explored the application of differential equations, specifically the Verhulst logistic equation, to the study of COVID-19.  By utilizing mathematical modeling and differential equations, we aimed to gain insights into the dynamics and trends of COVID-19 cases in Senegal over a certain time of period (April 2022 to April 2023).

First, we provided an overview of COVID-19, discussing its origins, transmission, and impact on
public health. We highlighted the importance of mathematical modeling in understanding the
spread and control of infectious diseases, particularly in the case of COVID-19.

Next, we introduced the Verhulst logistic equation, a widely used mathematical model for population growth. We adapted this model to analyze the spread of COVID-19, considering factors such as population size, total cases, total deaths, new cases , and new deaths. By fitting the Verhulst logistic equation to the COVID-19 data in Senegal, we were able to study the trend and dynamics of the disease over time.
Through our analysis, we observed the trend of COVID-19 cases in Senegal from April 2022
to April 2023. We examined the growth rate, peak periods, and fluctuations in case numbers,
providing valuable insights into the progression of the disease. We also compared two variations
of the logistic function: one without an additional constant term and the other with an additional
constant term. We demonstrated the relevance and effectiveness of applying differential equations,
specifically the Verhulst logistic equation, to the study of COVID-19. By analyzing the COVID-19 data in Senegal, we gained valuable insights into the trends, dynamics, and impact of the disease. Our analysis revealed that a shift of b in the parameter P, in the logistic function, provide a better fit for the Covid-19 data in Senegal. This research contributes to the broader field of epidemiology and provides valuable information for public health decision-making and future pandemic preparedness.
As further research and data become available, it is crucial to continue refining mathematical models, incorporating new variables, and considering the evolving nature of the pandemic. By combining rigorous mathematical modeling with empirical data, we can continue to enhance our understanding of COVID-19 and contribute to effective strategies for disease control and prevention.

\vspace{0.5cm}
\noindent {\bf Acknowledgments}: This work is done under support of AIMS Senegal. The authors would like to thank Dr. Bruno Sebastian Barton Singer for the comments and the suggestions.

\end{document}